\documentstyle[11pt]{article}
\begin{document}

\title{
\vspace*{-1cm}
\begin{flushright}
UR-1488
\end{flushright}
{\bf  Higher Dimensional SUSY Quantum Mechanics}}
\author{Ashok Das and Sergio A. Pernice \\
Department of Physics and Astronomy, \\
University of Rochester,\\
Rochester, New York, 14627. }
\maketitle

\begin{abstract}
Higher dimensional supersymmetric quantum mechanics is studied.
General properties of the two dimensional case are presented.
For three spatial dimesions or higher, a spin structure is
shown to arise naturally from the nonrelativistic supersymmetry
algebra.
\end{abstract}

The study of   supersymmetric quantum field theories in the low energy,
non-relativistic limit is of special interest for various reasons - primarily
because if supersymmetry is a symmetry of nature, what we see today
must be the low energy remnant of it. In such a limit, the underlying field
theory should approach a Galilean invariant  supersymmetric field theory
and, by the Bargmann super-selection rule~\cite{bar}, such a field theory
should be equivalent to a supersymmetric Schr\"{o}dinger equation in
each particle number sector of the theory. While one dimensional
supersymmetric quantum mechanics has been studied exhaustively in the
past~\cite{ew,coo}, there has only been a few attempts at generalizing
this to higher dimensions~\cite{cllo,crri,llm}. More importantly, to the best
of our knowledge, there has not been a systematic study of supersymmetric
quantum mechanics and its properties in higher dimensions. In this letter,
we propose a generalization of susy quantum mechanics to n-dimensions
which brings out a rich structure. Our results are quite general and will be
used in Ref.~\cite{dp}, in particular, to study the non-relativistic limit of a
$2+1$ dimensional supersymmetric field theory where dramatic
consequences regarding supersymmetry breaking arise.

Let us, very briefly, review one dimensional susy quantum mechanics for
completeness. Supersymmetric quantum mechanics is defined by the graded
algebra~\cite{ew}
\begin{eqnarray}\label{superalg1}
\left[ H , Q \right] = \left[ H , Q^\dagger \right] = 0 \quad, \\
\label{superalg2}
\{ Q , Q^\dagger \} = H \quad, \\
\label{superalg3}
\{ Q , Q \} = \{ Q^\dagger , Q^\dagger \} = 0.
\end{eqnarray}
where $Q$ and $Q^\dagger$ represent the supercharges while $H$ is the
dynamical Hamiltonian of the system.
Relation (\ref{superalg2}) indicates that the Hamiltonian $H$ can only
have positive or zero eigenvalues. Indeed, for an arbitrary state $| \psi >$,
according to (\ref{superalg2}), it must be true that,
\begin{eqnarray}\label{poseigenv}
< \psi | H | \psi > & = & < \psi | Q Q^\dagger | \psi > + 
< \psi | Q^\dagger Q | \psi >   \nonumber  \\
& = &  |Q^\dagger  | \psi >|^2   +  | Q | \psi >|^2 \ge 0.
\end{eqnarray}
If supersymmetry is not broken, that is, if the supercharges annihilate
the vacuum, then the ground state would have zero energy. On the other 
hand, if the supercharges do not annihilate the vacuum, then the vacuum
energy would be positive.

From (\ref{superalg3}), we see that the  supercharges  take the triangular
matrix form
\begin{equation}\label{superch}
Q = 
\pmatrix{
0 & 0 \cr
q & 0 \cr
}  \quad {\rm and} \quad 
Q^\dagger = 
\pmatrix{
0 & q^\dagger \cr
0 & 0 \cr
},
\end{equation}
where in 1 dimension, in units of $\hbar = 1$ and $m = 1/2$, one writes
\begin{equation}\label{Asusy}
q =  {d \over dx}  + W (x) \quad, \qquad
q^\dagger = - {d \over dx}  + W (x)
\end{equation}
$W(x)$ is known as the superpotential. The form  of the Hamiltonian,
then follows from~(\ref{superalg2}) and~(\ref{superch}) to be
\begin{equation}\label{susyhamil}
H = \pmatrix{
H_1 & 0 \cr
0 & H_2 \cr
} = \pmatrix{
q^\dagger q & 0 \cr
0 &  q q^\dagger \cr
}
\end{equation}
From the above equation we see that the ``superpartner" Hamiltonians 
have the form, in the coordinate basis,
\begin{equation}\label{suppartham}
H_1 = -  {d^{2} \over d x^2} + V_1 (x) \quad {\rm and} \quad
H_2 = -  {d^{2} \over d x^2} + V_2 (x),
\end{equation}
with
\begin{equation}\label{suppartpot}
V_1 (x) = W^2 (x) -  W' (x)  \quad {\rm and} \quad
V_2 (x) = W^2 (x) +  W' (x) .
\end{equation}
It is clear from (\ref{susyhamil}) that if  $| \psi >$ is an eigenstate of  
$H_1$ ($H_2$), then $q | \psi >$ ($q^\dagger | \psi >$), if different from 
zero, is an eigenstate of  $H_2$ ($H_1$) with equal energy. Superpartner 
states have equal energy.

Let us consider next, and study in some detail, the 2-dimensional case,
since it can be developed with standard technics and the results will be
used in ref.~\cite{dp}.  From eq.~(\ref{superalg2}) we see that a realization
of the supersymmetric algebra can be obtained, at the free level, only if the
Laplacian can be factorized into a pair of hermitian conjugate factors. In two
dimensions, the factorization appears naturally since we can write 
\begin{equation}\label{factor}
- \nabla^2 = \left( {\partial \over \partial x} + i {\partial \over \partial y} \right) 
\left( - {\partial \over \partial x} + i {\partial \over \partial y} \right)=
\left( -{\partial \over \partial x} + i {\partial \over \partial y} \right) 
\left(  {\partial \over \partial x} + i {\partial \over \partial y} \right)
\end{equation}
Then,  if  we define $q = \partial / \partial x + i \partial / \partial y$, both
$q q^\dagger$ and $q^\dagger q$ correspond to the negative of the
Laplacian, which would then describe the two dimensional free particle
supersymmetric quantum mechanics. In order to understand how to
construct a fully interacting two dimensional  supersymmetric theory
it is convenient to write $q$ in a frame independent way as follows
\begin{equation}\label{2-dqfree}
q_{\rm free} = \hat{e}^+ \cdot  \vec{\nabla}\ ,\quad  \hat{e}^+ = \hat{e}_x + i
\hat{e}_y
\end{equation}
were $\vec{\nabla}$ represents the two dimensional gradient. It is more or less
obvious now from~(\ref{Asusy}) and~(\ref{2-dqfree}) that the natural way to
introduce  interactions is through a {\it vector} super potential $\vec{W}$
such that
\begin{equation}\label{2-dq}
q = \hat{e}^+ \cdot  \left( \vec{\nabla} + \vec{W} \right) 
\end{equation}
From eq.~(\ref{2-dq}) we now get for the two superpartner Hamiltonians
\begin{eqnarray}\label{sh2d}
H_1 = q^\dagger q  =  \sum_{k = 1}^2 \left( -\nabla_k + W_k \right)
\left( \nabla_k + W_k \right) - i\epsilon^{ij} \left\{ \nabla_i , W_j \right\}  \\
H_2 = q q^\dagger  =  \sum_{k = 1}^2 \left( \nabla_k + W_k \right)
\left( -\nabla_k + W_k \right) - i\epsilon^{ij} \left\{ \nabla_i , W_j \right\}
\end{eqnarray}
were the curly brackets represent anti-commutators and $\epsilon^{ij}$
is anti-symmetric with $\epsilon^{12} = 1$. Note that the
vector super potential naturally generates a gauge field interaction
structure which results automatically from the supersymmetry algebra. We note
here that such a structure was, in fact, already noticed in the
non-relativistic limit of an interacting $2+1$ dimensional supersymmetric field
theory~\cite{llm}.

In polar coordinates, and assuming $W$ totally imaginary in analogy with
the gauge interactions, the supercharges take the form\footnote{ Note that in
d dimensions, the hermitian conjugate of the operator $\partial / \partial r$ is
 $-\partial / \partial r - \left( d-1 \right)/r$. In the reduced variables, which
we will use later, however,  $\left( \partial / \partial r \right)^\dagger = -
\partial / \partial r$.}
\begin{eqnarray}\label{sqpolar2d}
q &=& e^{i \theta}  \left[ {\partial \over \partial r} + {1 \over r}
i {\partial \over \partial \theta} + W(r,\theta) \right]   \\ \label{sqdaggpolar2d}
q^\dagger &=& \left[ - {\partial \over \partial r} + {1 \over r}
\left( i {\partial \over \partial \theta} -1 \right) + W^* (r,\theta) \right] e^{- i \theta} \\
\label{wrtheta}
W(r,\theta) &=& W_\theta  - i \ W_r 
\end{eqnarray} 
where in the last equation $W_r$ and $W_\theta$ refer to the radial and
the angular components of $\vec{W}$ respectively and a complex
$W(r,\theta)$ arises because of a complex projection.  We see from
eqs.~(\ref{sqpolar2d}) and~(\ref{sqdaggpolar2d})  that the two dimensional
supercharges, when acting on a given state, change its orbital  angular
momentum by one unit. As is shown in reference~\cite{dp}, this may have
dramatic consequences regarding supersymmetry breaking. We know
that in a supersymmetric field theory the supercharges change the total
angular momentum by half a unit. The two results are clearly perfectly
compatible if we keep in mind the spin degrees of freedom in a
supersymmetric field theory. Note that the ``superpotential" $W(r,\theta)$
is, in general, complex. We describe here the special case of a vector
potential $\vec{W} = - i \ W_{\theta} \ \hat{e}_{\theta}$, where the function
$W$ in~(\ref{wrtheta}) is real. The more general case can be worked out
just as easily.

For a rotationally invariant interaction, the two dimensional supersymmetric
formalism is reducible to the one dimensional formalism in the reduced
variables (i.e., on the space of wave functions $\psi = \sqrt{r}\varphi $, where
$\varphi$ is the total wave function). In this space, the supercharges become
(see footnote 1)
\begin{eqnarray}\label{sqpolar2dred}
q_{\rm red} &=& e^{i \theta}  \left[ {\partial \over \partial r} + {1 \over r}
\left( i {\partial \over \partial \theta}  - {1 \over 2}\right) + W_{\theta} \right]   \\ 
\label{sqdaggpolar2dred}
q_{\rm red}^\dagger &=& \left[ - {\partial \over \partial r} + {1 \over r}
\left( i {\partial \over \partial \theta} - {1 \over 2} \right) + 
W_{\theta} \right] e^{- i \theta}
\end{eqnarray} 
leading to the Hamiltonians
\begin{eqnarray}
H_{1, {\rm red}} & = & q_{\rm red}^{\dagger}q_{\rm red}\nonumber\\
 & = & \left[ - {\partial \over \partial r} + {1 \over r}
\left( i {\partial \over \partial \theta} - {1 \over 2} \right) + W_{\theta} \right]
\left[  {\partial \over \partial r} + {1 \over r}
\left( i {\partial \over \partial \theta} - {1 \over 2} \right) + W_{\theta}
\right]\nonumber\\\label{h1redext}
 & = & -{\partial^2\over \partial r^{2}} +  
{1 \over r^2}\left[ \left( i {\partial \over \partial \theta} \right)^2 - {1 \over 4} \right]
- {\partial W_{\theta} \over \partial r} + W_{\theta}^2  + {2 \over r} W_{\theta}   
\left( i {\partial \over \partial \theta}  - {1 \over 2} \right)  \\ 
H_{2, {\rm red}}  &=& q_{\rm red}q_{\rm red}^{\dagger}\nonumber\\
 &=&
\left[ {\partial \over \partial r} + {1 \over r}
\left( i {\partial \over \partial \theta} + {1 \over 2} \right) + W_{\theta} \right]
\left[ - {\partial \over \partial r} + {1 \over r}
\left( i {\partial \over \partial \theta} + {1 \over 2} \right) + W_{\theta}
\right]\nonumber\\\label{h2redext}
 &=&  - {\partial^{2} \over \partial r^{2}} +  
{1 \over r^2}\left[ \left( i {\partial \over \partial \theta} \right)^2 - {1 \over 4} \right]
+ {\partial W_{\theta} \over \partial r} + W_{\theta}^2  + {2 \over r} W_{\theta}   
\left( i {\partial \over \partial \theta}  + {1 \over 2} \right)
\end{eqnarray}
In eqs.~(\ref{h1redext}) and~(\ref{h2redext}) the first two terms correspond
to the free Hamiltonian in the reduced variables. The next two have the
structure of the one dimensional superpartner potentials of eq.~(\ref{suppartpot}).
The last term, a sort of topological (magnetic) angular momentum dependent
interaction, is present even for a spherically symmetric vector superpotential
as the one considered here (for a  vector superpotential which is not
spherically symmetric,
the corresponding term is $1/r  \left\{ i \partial / \partial \theta  - 1 / 2 , W_{\theta}
-  i W_r \right\}$). It arises automatically from the
two-dimensional  supersymmetry and, as we will show next, it ensures
the reduction of the two dimensional formalism to the one dimensional one
for a spherically symmetric vector superpotential in {\it each} angular momentum
sector independently. 

Let us now consider a state $| \psi_m >$ which is an eigenstate of  $H_1
= q_{\rm red}^\dagger q_{\rm red}$ with orbital angular momentum $m$.
Then, its superpartner is the state $q_{\rm red} | \psi_m >$, with orbital
angular momentum $m - 1$ which follows from (\ref{sqpolar2dred}). In this
space of states, the radial Hamiltonians take the form  
\begin{eqnarray}\label{radham2}
H_{1, {\rm red}}  &=&
\left[ - {\partial \over \partial r} + {1 \over r}
\left( m - {1 \over 2} \right) + W_{\theta} \right]
\left[  {\partial \over \partial r} + {1 \over r}
\left( m - {1 \over 2} \right) + W_{\theta} \right]  \\
H_{2, {\rm red}}  &=&
\left[ {\partial \over \partial r} + {1 \over r}
\left( m - {1 \over 2} \right) + W_{\theta} \right]
\left[ - {\partial \over \partial r} + {1 \over r}
\left( m - {1 \over 2} \right) + W_{\theta} \right]
\end{eqnarray}
We see then that the reduction, in a given angular momentum sector $m$,
corresponds to an effective one dimensional superpotential  of the form
($W_{\rm eff}$ is the same for both the superpartner states even though their
angular momenta are different)
\begin{equation}\label{weff}
W_{\rm eff} = {1 \over r} \left( m - {1 \over 2} \right) + W_{\theta}
\end{equation}
This, therefore, shows that in each angular momentum sector, the system
reduces to a one-dimensional susy system.
All of this can, of course, be checked explicitly in the simple example of an
isotropic oscillator which would correspond to a linear radial superpotential,
$W_{\theta} = a r$ were $a$ is a constant. However, this is straightforward and we
will not go into the details here.

Allowing, in general, the interaction to be dependent on $\theta$, we will
next find the consequences of two dimensional supersymmetry  on the
scattering amplitude for short ranged potentials ($W \rightarrow 0$ faster
than $1/r$). Let $| \psi >$ represent a scattering state of the Hamiltonian $H_1 = 
q^\dagger q$  corresponding to an incident plane wave along the $x$-axis,
then it will have the general asymptotic form
\begin{equation}\label{scatt}
< \vec{r} \ | \psi >  \ \stackrel{r \rightarrow  {\rm large}}{\longrightarrow} \ 
e^{i k r \cos{\theta}}  + {f_1  \left( \theta, k \right) \over  \sqrt{r}} e^{i k r}
\end{equation}
The superpartner state, on the other hand, is $q | \psi >$ and taking into
account the short range nature of the interaction, we obtain  from
eqs.~(\ref{sqpolar2d}) and~(\ref{scatt}),
\begin{equation}\label{scatt2}
< \vec{r} \ | \ q \ | \psi >  \ \stackrel{r \rightarrow  {\rm large}}{\longrightarrow} \
i k  \left[ e^{i k r \cos{\theta}}  + {f_1   \left( \theta, k \right) e^{i \theta}
 \over  \sqrt{r}} e^{i k r} \right]
\end{equation}
Comparing (\ref{scatt}) and (\ref{scatt2}), we, then, obtain
\begin{equation}\label{scatrel}
f_2  \left( \theta, k \right) = f_1   \left( \theta, k \right) e^{i \theta}
\end{equation}
which implies that, for short range interactions, the superpartner scattering
amplitudes differ only by a phase if supersymmetry is unbroken. Since 
\begin{equation}\label{scattprob}
|f_1 \left( \theta, k \right) |^2 = |f_2 \left( \theta, k \right) |^2
\end{equation}
the super partner states have identical probability for scattering by any
given angle.

We now proceed to generalize the above formalism to n-dimensions. As mentioned
earlier, even at the free level, a pre-condition to realize the supersymmetry
algebra of eqs.~(\ref{superalg1})-(\ref{superalg3}) is to be able to factorize
the Laplacian. More specifically, we need to find a $q$ such that $- {\nabla}^{2} = q
q^\dagger = q^\dagger q$. The problem is that in more than two dimensions
the Laplacian involves the sum of at least three terms, each one of which
is a second derivative. So it is clear that a  more general form for $q$, 
a linear combination of derivative operators with arbitrary complex 
coefficients, simply will not do the job. On the other hand, we also recognize
that we can realize the n-dimensional supersymmetry algebra easily if we
introduce noncommuting quantities.
Here the situation is very much like the case of the Dirac equation where one
can write the square root of the Einstein relation as a linear combination of
the energy-momentum operators provided one introduced non-commuting objects. 
Let us assume that $q$, in the free case, can be written as a linear
superposition
\begin{equation}\label{ndimq}
q = \sum_{j=1}^{n}{i_j \nabla_j}
\end{equation}
where the $i_j$'s are abstract objects which we assume to commute with 
$\nabla_j$'s . Then we have for $q q^\dagger$ and $q^\dagger q$
\begin{eqnarray}\label{nqqdagq}
q q^\dagger = \sum_{j=1}^{n}{i_j i_j^\dagger \  (-\nabla_j^2)} - \sum_{j  < k}^{n}
{\left( i_j i_k^\dagger + i_k i_j^\dagger \right) \  \nabla_j \nabla_k}   \\
q^\dagger q = \sum_{j=1}^{n}{i_j^\dagger i_j \  (-\nabla_j^2)} - \sum_{j < k}^{n}
{\left( i_j^\dagger i_k + i_k^\dagger i_j \right) \  \nabla_j \nabla_k}
\end{eqnarray}
Each of these expressions will equal the negative of the Laplacian  if the 
following relations are satisfied for every $j$ and $k$,
\begin{eqnarray}\label{relations1}
i_j i_j^\dagger &=& i_j^\dagger i_j =1\ ,\\   \label{relations2}
i_j i_k^\dagger + i_k i_j^\dagger &=& i_j^\dagger i_k + i_k^\dagger i_j = 0
\end{eqnarray}
The first equation allows for two possibilities, either  $i_j = i_j^\dagger$,
in which case $i_j^2 = 1$, or   $i_j = - i_j^\dagger$, which would imply
$i_j^2 = -1$. Some of the $i_j$'s may be hermitian while others may be
anti-hermitian. Thus, for example, in the two dimensional case discussed 
earlier, $i_1 =1$, while $i_2=i$. Furthermore, the second relation is a true
constraint only if the hermiticity properties are such that it is symmetric in
$j$ and $k$. In particular, we note that if all the $i_j$'s are Hermitian, the
eqs.~(\ref{relations1}) and~(\ref{relations2}) define a Clifford
algebra\footnote{We thank Prof. S. Okubo for pointing this out}.

Following our earlier discussion, we write $q$ in a frame independent
manner as
\begin{equation}\label{emaster}
q = \hat{\varepsilon} \cdot \vec{\nabla},\, \qquad\hat{\varepsilon} 
= \sum_{j=1}^n i_j \ \hat{e}_j
\end{equation}
This now allows us to introduce interactions, as before, through a
{\it vector} super potential $\vec{W}$ dependent on the position as
\begin{equation}\label{ndqint}
q = \hat{\varepsilon} \cdot \left( \vec{\nabla} +  \vec{W} \right) =
 \sum_{j=1}^n i_j  \left( \nabla_j +  W_j \right)
\end{equation}
It is now easy to calculate $q q^\dagger$ and $q^\dagger q$ using
eqs.~(\ref{relations1})-(\ref{relations2}) which take the forms
\begin{eqnarray}\label{ndqqdaggq1}
q q^\dagger = \sum_{j=1}^n \left( \nabla_j +  W_j \right) \left(-\nabla_j + W_j \right)
+ \sum_{j < k}  i_j i^\dagger_k \left[ \left\{ \nabla_j , W_k \right\}   -
\left\{ \nabla_k , W_j \right\} \right]   \\  \label{ndqqdaggq2}
q^\dagger q = \sum_{j=1}^n \left(-\nabla_j +  W_j \right) \left( \nabla_j + W_j \right)
- \sum_{j < k}  i^\dagger_j i_k \left[ \left\{ \nabla_j , W_k \right\}   -
\left\{ \nabla_k , W_j \right\} \right] 
\end{eqnarray}
We can see the emergence of a gauge interaction structure which
arises naturally from the requirement of supersymmetry. However, without
any additional specification of the algebra~(\ref{relations1})-
(\ref{relations2})  we can not go any further. It is
clear that one particular choice which would generalize our earlier
discussions of one and two dimensional susy quantum mechanics is
\begin{equation}
i_{1} = 1,\,\quad i_{j}^{\dagger} = - i_{j}\quad j\neq 1
\end{equation}
Another choice, for $n=3$, corresponds to the $i_j$'s  satisfying
 the {\it quaternionic} algebra
\begin{equation}\label{quat}
i_j i_k = - \delta_{j k} + \sum_{l=1}^3 \varepsilon_{jkl} \ i_l
\end{equation}
(where $\varepsilon_{jkl}$ is totally anti-symmetric with $\varepsilon_{123}=1$)
with the three $i_{j}$'s anti-hermitian
\begin{equation}\label{adjquat}
i^\dagger_j = - i_j
\end{equation}
It is interesting that many of the properties of quaternionic quantum
mechanics~\cite{adl} are also properties of supersymmetric quantum
mechanics. For example, in both of them the energy is positive semi-definite,
and a gauge interaction structure arises naturally.

We note that when
\begin{equation}\label{pauli}
i_j = \sigma_j
\end{equation}
$q$ has the form
\begin{equation}\label{spiq n}
q = \vec{\sigma} \cdot  \left( \vec{\nabla} + \vec{W} \right)
\end{equation}
suggesting a spin structure for the supersymmetric states. It is interesting to note
that while in 2 dimensions this was note necessary, a higher dimensional
susy quantum mechanics necessarily forces a spin structure into the theory.

We would like to thank Prof. S. Okubo for reading the manuscript and making
useful suggestions. This work was supported in part by the U.S. Dept. of
Energy Grant  DE-FG 02-91ER40685.

\end{document}